\bigskip
\documentclass[preprint,12pt]{elsarticle}
\usepackage{graphicx}
\usepackage{fancyhdr}
\usepackage{dcolumn}
\usepackage{amssymb}
\usepackage{amsmath}
\usepackage[normalem]{ulem}
\usepackage{bm}
\usepackage{textcomp}
  \usepackage{natbib}
  \usepackage[a4paper,left=1.55cm,top=2cm,right=1.55cm]{geometry}
\usepackage[colorlinks, linkcolor=red, anchorcolor=green,citecolor=blue ]{hyperref}
\usepackage[dvips]{color}

\journal{Physics Letters B}

\begin{document}

\begin{frontmatter}



\title{Level inversion in kaonic nuclei and the high-density nuclear equation of state}


\author{Rong-Yao Yang, Wei-Zhou Jiang, Si-Na Wei}

\address{School of Physics, Southeast University, Nanjing
211189, China}

\begin{abstract}
It is very difficult for any nuclear model to pin down the
saturation property and high-density equation of state (EOS)
simultaneously because of high nonlinearity of the nuclear
many-body problem. In this work, we propose, for the first time,
to use the special property of light kaonic nuclei to characterize
the relation between saturation property and high-density EOS.
With a series of relativistic mean-field models, this special
property is found to be the level inversion between orbitals
$2S_{1/2}$ and $1D_{5/2}$ in light kaonic nuclei. This level
inversion can serve as a theoretical laboratory to group the
incompressibility at saturation density and the EOS at
supra-normal densities simultaneously.
\end{abstract}


\end{frontmatter}

\section{Introduction}

The determination of the nuclear equation of state (EOS) with
certain accuracy is crucial for the understanding of the phenomena
related to relativistic heavy-ion collisions~\cite{Li08} and
astrophysical objects, such as, neutron
stars~\cite{98AA,Lattimer07}, gravity waves~\cite{Annala18}, etc.
Due to the high nonlinearity of the nuclear many-body problems,
some critical uncertainties still exist in the nuclear EOS after
decades of effort. For instance, the large uncertainty of the
high-density EOS in symmetric nuclear matter still has a large
relative error of 50\% as extracted from the data of the heavy-ion
collisions~\cite{02PD}. This large error not only prevents one
from extracting accurate structural information, such as the phase
structures and transitions, but also makes it more difficult to
pin down the huge divergency in symmetry energy at high
densities~\cite{Li08}. Though the high nuclear density can
naturally form in central neutron stars, neither can the celestial
data give a sufficiently satisfactory constraint on the
high-density EOS~\cite{Li08,Steiner13}.

Another important quantity of the nuclear EOS is the
incompressibility  at saturation
density~\cite{Stone14,RocaMaza18}. In the past, significant
progress has been achieved in constraining the incompressibility
by analyzing the experimental data, such as the nuclear masses,
nuclear radii and giant monopole
resonance~\cite{Blaizot80,Youngblood99,ToddRutel05,Klupfel09,Steiner10,Khan12,Stone14,Wang15,
RocaMaza18}, etc. The most proposed value for the
incompressibility is around 230 MeV, such as 230$\pm$40 MeV given
in Ref.~\cite{Khan12}, while higher values in the range of 250-315
MeV can not be ruled out after a comprehensive reanalysis of the
data on giant monopole resonances~\cite{Stone14}. Note that our
work does not aim  at narrowing down or excluding some range of
the incompressibility. Actually, this large uncertainty provides
us a parameter space to do the readjustment in the study of the
EOS.

Confronting the large uncertainty of the high-density EOS, one
would like to know whether there exist the sensitive probes, based
on the nuclear structure properties, to reduce it. On the other
hand, the high-density EOS should be connected to the EOS near or
beneath the saturation density, though the predictive power of
nuclear models is weak in the whole density region for the high
nonlinearity of the in-medium strong interaction.  One may ask
naturally: What is the relationship between the incompressibility
at saturation density and the EOS at high densities? It is also
very interesting to reveal what role of such a relationship can be
played in constraining the EOS at saturation and high densities
mutually. The key point to study these problems is firstly to have
a finite candidate with a high-density core that concerns the
high-density EOS.

In fact, the kaonic nuclei are a good candidate  that can feature
a dense core at supra-normal density~\cite{02TY,02YA}. It is
theoretically feasible to employ structural properties of kaonic
nuclei that encompass the information of a large density range to
impose a global constraint on the EOS both in the lower-density
and higher-density regions.  Compared with the data at high
densities extracted from heavy-ion collisions,  the structural
properties in the ground state of kaonic nuclei should be more
precise for probing the high-density EOS. In the past, we have
actually made some efforts to establish the correlation between
the low-density halo phenomenon in kaonic nuclei and the stiffness
of EOS at saturation and high densities~\cite{Yang17}. In this
work, we further investigate the ground-state properties of kaonic
nuclei in the framework of relativistic mean-field (RMF) theory.
It is one of our aims to explore the relationship between the
incompressibility and the stiffness of the high-density EOS by
singling out some special structural information in kaonic nuclei.
We will find that the level inversion characteristic in light
kaonic nuclei is tightly related to incompressibility,  the
stiffness of the high-density EOS,  and the relationship between.
In the following, we in turn give the formalism, results and
discussions, and the summary in sect. II-IV.

\section{Formalism}

In kaonic nuclei, the binding of the antikaon is given by its
additive couplings with scalar and vector mesons, whereas this is
apart from the nucleonic binding in normal nuclei that is a
cancellation of the scalar and vector potentials. Such a deep
binding in kaonic nuclei results in correspondingly much deeper
nucleonic potential due to the coupling between the antikaon and
nucleons, giving rise to exotic features such as the dense core at
supra-normal density~\cite{02TY,02YA} and diffusive
halo~\cite{Yang17}. Here, we adopt the RMF theory to investigate
the ground-state  properties of kaonic nuclei and its relationship
with the nuclear EOS. The Lagrangian density for nucleons is
generally given by
\begin{eqnarray} \label{LN}
\mathcal{L}_0 &=&  \bar{\psi}_B [ i\gamma_{\mu} \partial ^{\mu} - M_{B} + g_{\sigma} \sigma - g_{\omega} \gamma_{\mu}
\omega^{\mu} - g_{\rho} \gamma_{\mu} \tau_3 b_0^{\mu}
- e \frac{1+\tau_3}{2} \gamma_{\mu} A^{\mu}] \psi_B - \frac{1}{4}
F_{\mu \nu} F^{\mu \nu}+\frac{1}{2} m^2_{\omega} \omega_{\mu} \omega^{\mu} \nonumber  \\
& & - \frac{1}{4} B_{\mu \nu} B^{\mu \nu} + \frac{1}{2}m^2_{\rho} b_{0\mu} b_0^{\mu}-
\frac{1}{4} A_{\mu \nu} A^{\mu \nu}
+ \frac{1}{2} (\partial _{\mu}\sigma \partial ^{\mu} \sigma - m^2_{\sigma} \sigma^2) + U(\sigma,\omega^\mu),
\end{eqnarray}
where $\psi_B$,  $\sigma,\ \omega_{\mu},\ b_{0\mu},\ A_{\mu}$
represent the baryon, scalar, vector, isovector vector,
electromagnetic field, respectively. The $g_i$
($i=\sigma,\omega,\rho$) is the corresponding coupling constants
between meson fields and nucleons. The $M_B$, $m_{\sigma}$,
$m_{\omega}$, $m_{\rho}$ are the masses for baryon and
corresponding mesons. The $\tau_3=\pm1$ for protons and neutrons,
respectively. The strength tensors of the $\omega$, $\rho$ meson
and the photon are defined as $F_{\mu
\nu}=\partial_{\mu}\omega_{\nu}-\partial_{\nu}\omega_{\mu}, B_{\mu
\nu}=\partial_{\mu}b_{0\nu}-\partial_{\nu}b_{0\mu}, A_{\mu
\nu}=\partial_{\mu}A_{\nu}-\partial_{\nu}A_{\mu}.$

The self-interacting terms are expressed as
\begin{eqnarray}
U(\sigma,\omega^\mu)&=& - \frac{1}{3} g_2\sigma^3 - \frac{1}{4} g_3 \sigma^4 +
\frac{1}{4} c_3 (\omega_\mu \omega^\mu)^2,
\end{eqnarray}
with $g_2,\ g_3$ denoting the strength of the  nonlinear terms of
scalar field. The $c_3$ is the parameter for the non-linear
$\omega$ self-coupling term with which the stiffness of the
nuclear EOS  at high densities can be readjusted sensitively. In
the RMF approximation~\cite{Walecka86}, it is easy to deduce the
EOS for nuclear matter and the equation of motion for nucleons in
a nucleus. The sound velocity square $v_s^2$ is defined as
$v_s^2=\partial P/\partial\epsilon$ being the partial derivative
of the pressure with respect to the energy density, and it can be
used to measure the stiffness of the nuclear EOS.

To investigate kaonic nuclei, we should incorporate  the
Lagrangian density describing the strong attraction between the
nucleons and $K^-$ mesons. The Lagrangian density for kaons is
written as~\cite{07DG}
\begin{eqnarray}
\label{LK}
\mathcal{L}_{KN} = (\mathcal{D}_{\mu}K)^{\dag}(\mathcal{D}^{\mu}K)
 - (m^2_{K} - g_{\sigma K} m_{K} \sigma )K^{\dag}K,
\end{eqnarray}
where the covariant derivative is defined as $\mathcal{D}_{\mu}
\equiv \partial _{\mu} + i V_\mu$ with $V_\mu = g_{\omega K}
\omega_{\mu} +  g_{\rho K} b_{0\mu} + e\frac{1+\tau_3}{2}A_{\mu}$.
Here, $K=K^+$ and  $K^\dag=K^-$ instead of kaonic doublets since
in kaonic nuclei it is just a $K^-$ meson implanted into a nucleus
without neutral kaons. It can be inferred from Eq.(\ref{LK}) that
the strongly attractive $K^-N$ interaction is mediated by
$\sigma,\ \omega,\ \rho$ meson fields and electromagnetic field.
This strong attraction is responsible for the deep binding of the
kaonic nuclei.

Combining the Lagrangian density for nucleons in Eq.~(\ref{LN})
and $K^-$ mesons in Eq.~(\ref{LK}), we can write out the RMF
equations of motion as follows
\begin{eqnarray}
[-i \bm{\alpha} \cdot \bm{\nabla} + \beta ( M_{B} - g_{\sigma}
\sigma ) + g_{\omega} \omega +
 g_{\rho} \tau_3 b_0
   +e\frac{1+\tau_3}{2} A]\psi_B &=& E \psi_B, \label{EqD} \\
( -\bm{\nabla}^2 +  m^2_{K} -E^2_{K^-} + \Pi )K^- &=&0,
\label{KG_K} \\
 ( -\bm{\nabla}^2 + m^2_\phi )\phi &=& s_\phi,
\end{eqnarray}
where the $s_\phi$ denotes the source term for mesons or photon
($m_{A}=0$)
\begin{eqnarray}
s_\phi =\left\{
\begin{aligned}{}
&g_{\sigma} \rho_s - g_2\sigma^2 - g_3\sigma^3 + g_{\sigma K}m_K K^-K^+,  & \phi=\sigma, &\\
&g_{\omega} \rho_v  -c_3\omega^3 - g_{\omega K}\rho_{K^-}, & \phi=\omega, &\\
&g_{\rho} \rho_3 - g_{\rho K}\rho_{K^-}, & \phi=b_0, &\\
&e \rho_p - e\rho_{K^-}, & \phi=A. &\\
\end{aligned}
\right.
\end{eqnarray}
Here $\rho_s,\ \rho_v,\ \rho_3,\ \rho_p$ are the
scalar, vector, isovector, and proton density, respectively.
$\rho_{K^-}$ is the $K^-$ density~\cite{14RYY}
\begin{equation}
\rho_{K^-}=2(E_{K^-}+g_{\omega K}\omega+g_{\rho K}
b_0+eA)K^-K^+.
\end{equation}
$E$ and $E_{K^-}$ are the
energy eigenvalues for nucleons and $K^-$ meson, respectively. The
$K^-$ self-energy term is expressed as
\begin{equation}
\label{SK}
\Pi = - g_{\sigma K} m_{K} \sigma  - 2E_{K^-}V - V^2,
\end{equation}
where $V = g_{\omega K} \omega +  g_{\rho K}b_{0} +
\frac{1+\tau_3}{2}eA$. The parameters for $K^-$ are adapted to the
$K^-$ optical potential which is defined as
$U_{opt}(\textbf{p},\rho_0)
=\omega(\textbf{p},\rho_0)-\sqrt{\textbf{p}^2 + m_{K}^2}$ with
$\omega(\textbf{p},\rho_0)$  being the in-medium energy at saturation
density~\cite{14ZQF}.

Solving these equations of motion with an iterative method, one
can  calculate the ground-state properties of normal nuclei and
kaonic nuclei after the parametrizations are chosen.  More
detailed information can be referred to  Ref.~\cite{14RYY}.

\section{Results and discussions}
Prior to giving the numerical results, we first elaborate the
parametrization. For the nucleonic sector, we take some best-fit
non-linear RMF models that fit the saturation properties of
symmetric nuclear matter and the ground-state properties of
typical spherical nuclei. For the kaonic sector, the determination
of the parameters $g_{\omega K}$ and $g_{\rho K}$ in
Eq.~(\ref{LK}) can resort to the SU(3) relation: $2g_{\omega
K}=2g_ { \rho K}=g_ { \rho \pi}=6.04$, and the $g_{\sigma K}$ is
adjusted to the $K^-$ optical potential.  Up to now, there is no
consensus on the depth of $K^-$ optical potential. Experimentally,
it can be extracted from the data of kaonic
atoms~\cite{81batty,94EF,97CJB,99EF,01gal,07EF},  $K^-N$
scatterings~\cite{96waas,97waas,97JSB,00AR,01AC,11AC} and
heavy-ion collisions~\cite{14ZQF,97GQL,99WC,99FL,03AF,06WS}.
Because the data from the $K^-N$ scatterings and kaonic  atoms
mainly reflect the low-density feature of $K^-$N interaction, one
needs some specific extrapolations to get the depth at saturation
density. Unfortunately, the extrapolated values are largely model
dependent, ranging from -200 MeV to -50
MeV~\cite{81batty,94EF,97CJB,99EF,01gal,07EF,96waas,97waas,97JSB,00AR,01AC,11AC}.
On the other hand, the heavy-ion collisions can   produce an
almost consistent $K^-$ potential depth around -100 MeV at
saturation density~\cite{14ZQF,97GQL,99WC,99FL,03AF,06WS}. Hence,
we adopt -100 MeV as the $K^-$ optical potential depth at
saturation density~\cite{14ZQF} to fit the kaonic parameters. We
should mention that the effects from the $K^-$ absorption by
nucleons can be included by introducing an imaginary potential
into the $K^-$ self-energy term $\Pi$ in
Eq.~(\ref{SK})~\cite{06JM,06XHZ,07DG}. It is found that the
inclusion of the imaginary potential has little effect on the
final results  in our parametrizations and does not ever affect
the conclusion of this work. Therefore, we will not deal with the
imaginary part in detail hereafter.

\begin{figure}[thb]
\centering
\includegraphics[width=6cm]{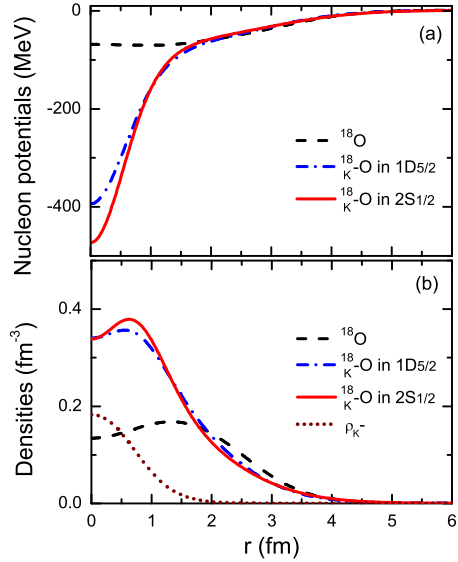}
\caption{(Color online) The nucleon potentials and nuclear
densities in $^{18}O$ (dashed curves) and $^{18}_{K^-}O$
(dash-dotted and solid curves). The labels "$1D_{5/2}$" and
"$2S_{1/2}$" denote the out-most two neutrons occupying
orbitals $1D_{5/2}$  and $2S_{1/2}$,  respectively. In the
lower panel, the $K^-$ density distribution $\rho_{K^-}$ is
obtained with the out-most neutrons occupying orbital $2S_{1/2}$. }
\label{Fpot}
\end{figure}

Firstly, we adopt NL3 parameter~\cite{NL3} for nucleons in
Eq.~(\ref{LN}) to explore the  ground-state properties in kaonic
nuclei. Displayed in Fig.~\ref{Fpot} are the nucleon potentials
and nuclear densities in $^{18}O$ and $^{18}_{K^-}O$. Due to the
strong attraction between nucleons and $K^-$ mesons, the nucleon
potential in kaonic nuclei is greatly changed by the implantation
of the $K^-$ meson.  We can see that the nucleon potential in
kaonic nucleus $^{18}_{K^-}O$ is much deeper than that in normal
nucleus $^{18}O$ in the core region ($r<2fm$). Its direct
consequence is that the nuclear density in the core of
$^{18}_{K^-}O$ is over two times the saturation density,
dramatically larger than that of $^{18}O$. We notice that the
implanted $K^-$ meson is almost squeezed in the core region where
the huge variation in the nucleon potential takes place
consistently, as seen in Fig.~\ref{Fpot}. For comparison, we
present  in Fig.~\ref{Fpot} the results for two orbital occupation
orders in $^{18}_{K^-}O$: one with the out-most two neutrons
occupying the orbital $1D_{5/2}$  and the other occupying orbital
$2S_{1/2}$. As can be seen, the latter corresponds to a higher
nuclear density and a deeper nucleon potential in the core region
which may imply that the occupation order with the out-most
orbital $2S_{1/2}$ is energetically favorable. It is checked
explicitly that the total binding energy of $^{18}_{K^-}O$ with
the out-most orbital being  $2S_{1/2}$ is truly larger than that
with the out-most orbital being $1D_{5/2}$. In this case,  the
ground state of $^{18}_{K^-}O$ is affirmed to have the occupation
order whose out-most orbital is the $2S_{1/2}$.  Together with the
internal level shifts, the level ordering appears to be different
from Mayer's shell model.

\begin{figure}[thb]
\centering
\includegraphics[width=8cm]{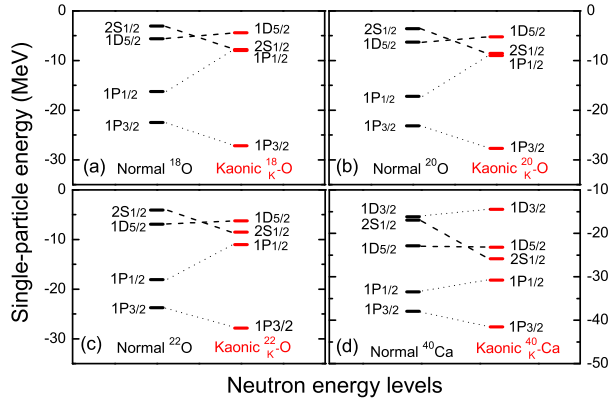}
\caption{(Color online) The energy levels for neutrons in
Oxygen isotopes, $^{40}$Ca and their corresponding kaonic nuclei.}
\label{Fspec}
\end{figure}

In Fig.~\ref{Fspec}, we show the energy levels for neutrons in
Oxygen isotopes, $^{40}$Ca and their corresponding kaonic nuclei.
Indeed, the orders of energy levels in all these kaonic nuclei are
not the same as Mayer's shell model: the orbital $2S_{1/2}$
obviously bounds deeper than the orbital $1D_{5/2}$. It appears
that orbitals $1P_{1/2}$, $2S_{1/2}$ and $1D_{5/2}$  form a new
shell in kaonic nuclei. In these cases, the magic number is
simultaneously changed. We also calculated the ground-state
properties for $^{32}_{K^-}$S, $^{42}_{K^-}$Ca, $^{44}_{K^-}$Ca,
etc., and found that the downward shift of $S_{1/2}$ orbitals is
very general in kaonic nuclei. As this level-inversion phenomenon
does exist pervasively  in light and intermediate-light kaonic
nuclei, it is necessary to reveal the underlying mechanism
responsible for the very phenomenon and the corresponding spectra
specific for kaonic nuclei.

The phenomenon of the downward shift of $S_{1/2}$  orbitals can be
understood from the corresponding schr\"{o}dinger-like equation
for the big component of nucleon's Dirac equation Eq.~(\ref{EqD}).
To solve the Dirac equation, we introduce the quantum number set
for a spherical nucleus which is $\{n,\kappa,m, t\}$ where
\emph{n} is the principle quantum number, \emph{m} is the magnetic
quantum number, and $t$ denotes the isospin. $\kappa$ is the
eigenvalue of a conserved operator  \textbf{K} with
$\mathbf{K}\equiv -\gamma^0 (\mathbf{\Sigma \cdot L}+1)$,
measuring the spin-orbit coupling. With these quantum numbers, the
four-component Dirac spinor can be divided into two components,
\begin{equation}\label{GF}
\psi_{n\kappa m t}(r)=\left\{\begin{array}{c} i\frac{G_a(r)}{r}\Phi_{\kappa,m}\\
-\frac{F_a(r)}{r}\Phi_{-\kappa,m}\end{array} \right\} \chi_t,
\end{equation}
where $G_a(r)$ and $F_a(r)$ are the big and small components of
the (radial) spinor, respectively. $\{a\}=\{n,\kappa,t\}$, $\Phi$
is the  spinor spherical harmonic, and $\chi_t$ is the isospinor
with $t=\pm1$ for protons and neutrons, respectively. Substituting
Eq.~(\ref{GF}) into the Dirac equation Eq.~(\ref{EqD}), the
corresponding Schr\"{o}dinger-like equations for the big component
can be obtained as
\begin{eqnarray}
\label{eqG}
[\frac{d^2}{dr^2}-\frac{\kappa(\kappa+1)}{r^2}+\frac{1}{U_G}\frac{d\triangle}{dr}\frac{d}{dr}
+\frac{1}{U_G}\frac{d\triangle}{dr}\frac{\kappa}{r} -
 U_GU_F ]G_a(r) = 0,
\end{eqnarray}
where $U_G=M_B+E-\triangle$,  $U_F=M_B-E+\Sigma$, and $\Sigma =
V(r)+S(r), \hbox{ }\triangle=V(r)-S(r)$ with $V(r)=g_{\omega}\omega+g_{\rho}\tau_3 b_0+eA(1+\tau_3)/2$, and
$S(r)=-g_{\sigma}\sigma$. As the $\kappa=-1$ for $S_{1/2}$
orbitals, the repulsive centrifugal  barrier (CB) vanishes as
$\kappa(\kappa+1)/r^2$=0. Then, there is no repulsive force
contributed from CB to counteract the strongly attractive
interaction induced by $K^-$ meson in the core region of kaonic
nuclei. Naturally, the $S_{1/2}$ orbitals are driven greatly
downward by the implanted $K^-$ meson. The level inversion may
occur for enough downwards shift of the $S_{1/2}$ orbital. For
instance, the level inversion between orbitals $2S_{1/2}$ and
$1D_{5/2}$ is very prominent in light kaonic nuclei. We notice
that the downward shift of $S_{1/2}$ orbitals becomes shallower
for heavier kaonic nuclei since the $K^-$ meson will be pulled
outwards by the strong attraction provided by more and more
out-layer nucleons. Consistently, a shallower nucleon potential
appears near the center of heavier kaonic nuclei. Consequently,
the $2S_{1/2}-1D_{5/2}$ inversion diminishes in intermediate-mass
kaonic nuclei and will vanish in heavy kaonic nuclei.

From Fig.~\ref{Fspec}, we can also observe that the spin-orbit
splitting of orbitals $1P_{3/2}$ and $1P_{1/2}$  in kaonic nuclei
is obviously larger than that in corresponding normal nuclei. This
is attributed to the enhanced spin-orbit potential due to the
implantation of the $K^-$ meson, while the latter provides an
additional strong attraction to deepen the nucleon potential and
hence increase  the radial gradient of the potential that accounts
for the  spin-orbit potential. Consequently, it makes the
spin-orbit splitting of the $1P_{3/2}$ and $1P_{1/2}$ orbitals in
kaonic nuclei much larger than that of normal nuclei. Here, we
stress that the downward shift of $S_{1/2}$ orbitals dominates the
level re-orderings including the level inversion and the change of
the spin-orbit splittings. The similar role of $2S_{1/2}$ orbital
in influencing the energy levels was pointed out in
Refs~\cite{04Rutel,06Jiang}. Especially, it was found  that the
occupation of $2S_{1/2}$ proton orbital plays  an important role
in enhancing  the $P_{3/2}-P_{1/2}$ splitting of neutron orbitals
in $^{48}$Ca as compared to $^{46}$Ar~\cite{04Rutel}. We should,
however, note that the underlying mechanism and the consequence
for the shift of $S_{1/2}$ orbitals are quite different in this
work. The large shift here does change the normal order of energy
levels, while it does not in the normal nuclei.  More importantly,
the occupation of $S_{1/2}$ orbitals in normal nuclei has nothing
to with the high-density core that is only subject to the
additional strong attraction provided by the $K^-$ implantation.
The large shift  of $S_{1/2}$ orbitals pulled down by $K^-$ meson
actually reflects a result of the compression that may give rise
to a core density up to twice the saturation density. The kaonic
nuclei can thus be a possible platform to reveal the effects of
different high-density EOS's by virtue of theoretical signals such
as the downward shift of $S_{1/2}$ orbitals, the broadened
spin-orbit splitting of $1P$ orbitals, and other structural
changes  in kaonic nuclei.

The effects of high-density EOS can be reflected  in the
structural characteristic of kaonic nuclei since the
aforementioned downward shift of $S_{1/2}$ orbitals could produce
supra-normal nuclear density near the center. Additionally, the
incompressibility should also be an important factor which would
have considerable impact on the nucleon potential. It is thus very
meaningful to study the relationships between the ground-state
properties of kaonic nuclei and the incompressibility and
high-density stiffness of the nuclear EOS. In the following, we
will firstly obtain a series of EOS's with different
incompressibility and stiffness at high densities. Then, the
corresponding parameters are used  to calculate the ground-state
properties of kaonic nuclei. To simulate different EOS's
systematically, we set the masses in Eq.(\ref{LN}) the same as
that of NL3~\cite{NL3}. With a given $c_3$, the coupling constants
including $g_\sigma,\ g_\omega,\ g_2,\ g_3$ are fitted to the
saturation properties, including the energy per baryon -16.3 MeV
at saturation density $\rho_0=0.148fm^{-3}$, the effective mass of
nucleon $M_B^{\ast}/M_B$ around 0.6 and a given incompressibility
(adjustable). Generally, a larger $c_3$ would produce a softer
EOS. These simple treatments could produce a series of EOS's with
the different incompressibility or different stiffness (sound
velocity square) at high densities. Though these parameter sets
are not obtained by the best-fit procedure,   the good accuracy in
simulating  the properties of finite nuclei is comparable to those
best-fit parameter sets. At the same time, the main conclusion in
this paper is not affected by the strategy of choosing the base
EOS. Based on these obtained nuclear EOS's, we can calculate the
ground-state properties of kaonic nuclei and its relationship with
the nuclear EOS. By doing so, a clear connection can be built
between the nuclear EOS and the phenomenon of level inversion in
kaonic nuclei.


\begin{figure}[thb]
\centering
\includegraphics[width=6cm]{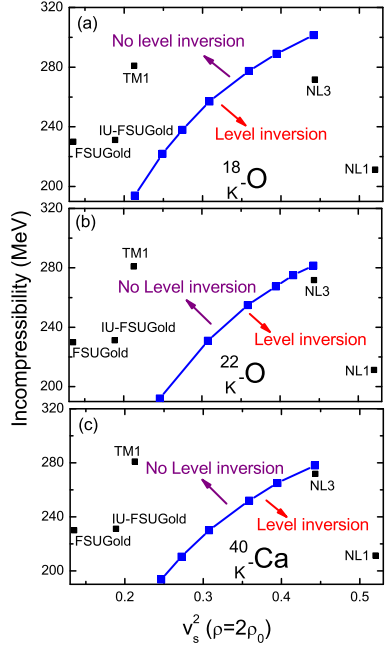}
\caption{(Color online) The phase diagram for the level inversion
in $^{18}_{K^-}O$, $^{22}_{K^-}O$ and $^{40}_{K^-}Ca$ in the
incompressibility-stiffness plane.  Here, we use the sound
velocity square (the partial derivative of the pressure with
respect to the energy density) at 2$\rho_0$ as a criterion of the
EOS's stiffness at high density. } \label{Finver}
\end{figure}

Using $^{18}_{K^-}$O, $^{22}_{K^-}$O and $^{40}_{K^-}$Ca as
examples, we show in Fig.~\ref{Finver} the critical level
inversion in the incompressibility-stiffness plane. It is found
that there exists critical boundary to separate the system with and
without the level inversion.  As can be seen from
Fig.~\ref{Finver}, the level inversion will occur for a stiff EOS
at high densities combined with a relative small incompressibility
(on the lower right side of the critical boundary in each panel),
while it does not occur for  a relative soft EOS at high densities
combined with a large incompressibility (on the upper left side of
the boundary). This result is understandable since the level
inversion is dominated by the downward shift  of $S_{1/2}$
orbitals which is tightly related to considerable compression near
the center of kaonic nuclei induced by the strong $K^-$N
attraction. A smaller incompressibility means easier  compression,
and a stiffer EOS corresponds to a deeper $K^-N$ attraction at
high densities as the interactions mediated by scalar and vector
mesons are coherently attractive. For the EOS's above critical
boundary, the downward shift  of $S_{1/2}$ orbitals remains, but
the $2S_{1/2}$ orbital becomes less  bound  than the $1D_{5/2}$
orbital.

According to the incompressibility and the 2$\rho_0$ sound
velocity square of the model, we mark in Fig.~\ref{Finver} the
positions for several best-fit parameter sets, such as
FSUGold~\cite{FSUG}, IU-FSUGold~\cite{IUFSUG}, NL1~\cite{NL1},
TM1~\cite{TM1}, etc. The allocation of these parameter sets on the
both sides of the critical boundary is consistent with their
predication on the situation of the $2S_{1/2}-1D_{5/2}$ inversion
in $^{18}_{K^-}$O, $^{22}_{K^-}$O and $^{40}_{K^-}$Ca. Namely, the
level-inversion phenomenon appears with the parameter sets NL1 and
NL3, while it does not appear with the FSUGold, IU-FSUGold, and
TM1.  Hence, nuclear EOS's can be divided into different groups by
this critical boundary in the incompressibility-stiffness plane.
As seen in Fig.~\ref{Finver}, the critical boundaries obtained
from different kaonic nuclei are similar with each other, whereas
the differences between each curve are subject to  the difference
of nucleon potentials in $^{18}$O, $^{22}$O and $^{40}$Ca. The
critical boundary can be shifted moderately when the parameter
sets are refitted to a different nucleon effective mass that may
affect the density of the energy level. The shift may also be
caused by the corresponding modification of the nucleon potential.
Nevertheless, these critical boundaries  can give the same
grouping for different EOS's. Here, we can see that the level
inversion  needs a stiffer EOS at high densities  combined
necessarily with a relatively smaller incompressibility, while a
softer EOS at high densities with larger incompressibility
declines the occurrence of the level inversion. If the level
inversion does not occur, the natural combination should be a
stiffer EOS at high densities with larger incompressibility or a
softer EOS with smaller incompressibility.

\begin{figure}[thb]
\centering
\includegraphics[width=8cm]{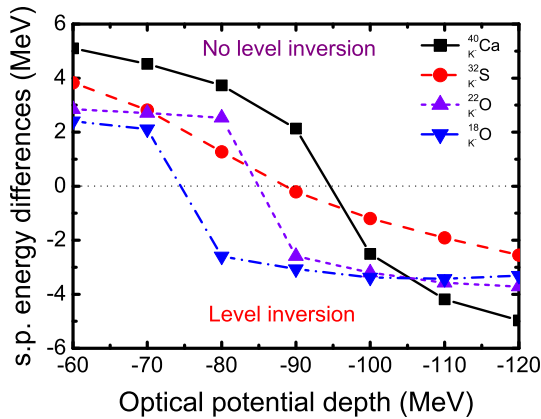}
\caption{(Color online) The single-particle energy differences
between $2S_{1/2}$ and $1D_{5/2}$
orbitals as a function of the depth of $K^-$ optical
potential at saturation for  $^{40}_{K^-}$Ca, $^{32}_{K^-}$S,
$^{22}_{K^-}$O, and $^{18}_{K^-}$O. Here, the RMF
parameter set NL3 is adopted in the calculation. } \label{FBE}
\end{figure}

To investigate the influence of the $K^-$ optical potential depth,
we display in Fig.~\ref{FBE} the single-particle (s.p.) energy
differences between $2S_{1/2}$ and $1D_{5/2}$ orbitals as a
function of the depth of $K^-$ optical potential at saturation for
$^{40}_{K^-}$Ca, $^{32}_{K^-}$S, $^{22}_{K^-}$O, and
$^{18}_{K^-}$O. The level inversion requires a deep potential
provided by the attraction of $K^-$. As clearly seen in
Fig.~\ref{FBE}, there are large parameter spaces especially for
light kaonic nuclei allowing the occurrence of the level
inversion. In fact, the level inversion will occur inevitably
after the $K^-$ optical potential exceeds the threshold. This
threshold becomes larger for heavier nuclei, as this is a result
of the dilution of the $K^-$ strong attraction by more and more
nucleons. For light kaonic nuclei, the threshold can be smaller
significantly. For a particular kaonic nucleus, the critical
boundary  in Fig.~\ref{Finver} will shift moderately to the upper
left (lower right) side for deeper (shallower) $K^-$ optical
potential. Interestingly, we also find that kaonic nuclei as light
as $^{14}_{K^-}$C have a similarly large parameter space for the
$2S_{1/2}-1P_{1/2}$ inversion. Accordingly, a large parameter
space including the atomic number allows one to choose appropriate
kaonic nuclei with the manifest level inversion to group different
EOS's, though the depth of the $K^-$ optical potential should be
unique and  eventually extracted from data.

Knowing that kaonic nuclei are yet to be verified experimentally,
we have focused in this work  on the special properties of kaonic
nuclei and their implication to the EOS especially at high
densities. These theoretical results suggest that the experiments
to  either identify or exclude the kaonic nuclei are very
necessary and valuable. We hope the further
experiments~\cite{17Nagae,19JPARC} could bring us affirmative
information on this subject in the future.

\section{Summary}
We have investigated the ground-state properties in kaonic nuclei
and its relationship with the nuclear EOS in the framework of RMF
theory. Owing to the strong attraction between nucleons and $K^-$
meson, supra-normal nuclear density appears near the center of
kaonic nuclei, and  the structure of energy levels in kaonic
nuclei becomes different from Mayer's shell model. The most
striking feature is that the $S_{1/2}$ orbitals shift downwards
significantly and the inversion between $2S_{1/2}$ and $1D_{5/2}$
orbital  may occur. This level inversion phenomenon is intimately
related to the incompressibility and the stiffness of the nuclear
EOS at high densities. Specifically,  according to the level
inversion  in kaonic nuclei, there exists a critical boundary  in
the incompressibility-stiffness plane   to divide the nuclear EOS
into different groups. The normal level order without the
inversion supports a natural combination: the stiffer EOS at high
densities with larger incompressibility or the softer EOS with
smaller incompressibility, while the occurrence of the level
inversion agrees with the combination of stiffer EOS at high
densities and smaller incompressibility. These findings provide us
theoretical tools to understand/constrain the nuclear EOS in the
whole density region.

\section*{Acknowledgements}
The work was supported in part by the National Natural Science
Foundation of China under Grant Nos. 11775049 and 11275048.

\end{document}